\documentclass[twocolumn,showpacs,aps,prl,superscriptaddress]{revtex4}

\usepackage{graphicx}
\usepackage{dcolumn}
\usepackage{amsmath}

\begin{document}

\affiliation{Brookhaven National Laboratory, Upton, New York 11973}%
\affiliation{University of Science \& Technology of China, Anhui 230027, China}%

\author{P.~Sorensen}\affiliation{Brookhaven National Laboratory, Upton, New York 11973}
\author{X.~Dong}\affiliation{University of Science \& Technology of China, Anhui 230027, China}

\title{ \vspace*{-0.5cm} Suppression of Non-photonic Electrons from
  Enhancement of Charm Baryons in Heavy Ion Collisions}\noaffiliation

\date{\today}

\begin{abstract}

At intermediate transverse momentum ($2<p_T<6$~GeV/c), the
baryon-to-meson ratio in Au+Au collisions is enhanced compared to p+p
collisions. Since charm-baryon decays produce electrons less
frequently than charm-meson decays, the non-photonic electron spectrum
is sensitive to the $\Lambda_c/D$ ratio. In this report we study the
dependence of the non-photonic electron spectrum on the
baryon-to-meson ratio for charm hadrons.  As an example, we take the
$\Lambda_c/D$ ratio to have the same form as the $\Lambda/K^0_S$
ratio. In this case, even if the total charm quark yield in Au+Au
collisions scales with the number of binary nucleon-nucleon collisions
($N_{bin}$), the electron spectrum at $2<p_T<5$~GeV/c is suppressed
relative to $N_{bin}$ scaled p+p collisions by as much as 20\%.

\end{abstract}


\maketitle

\vspace{0.5cm}

{\it Introduction.} --- Non-photonic electrons from heavy flavor
decays can be used to study charm production even when direct
measurements of heavy flavor hadrons are experimentally
unfeasable. Radiative energy loss models that successfully describe
the large hadron suppression in central Au+Au collisions 
predict a smaller energy loss for heavy flavor quarks (the dead-cone
effect)~\cite{deadcone}. Recent measurements, however, show that for
$3<p_T<8$~GeV/c the non-photonic electron spectrum in central Au+Au
collisions is supressed by a factor of five compared to expectations
from $N_{bin}$ scaled p+p collisions: a suppression that is as large
as that seen for charged hadrons~\cite{phelectron,stelectron}. No
radiative energy loss models predict such a large suppression for
heavy-flavor quarks. Facing the large discrepancy between the parton
radiative energy loss results and data, scenarios that involve
collisional energy loss are being revisited~\cite{collisional}. In
addition, we believe that this discrepency may be partially resolved by
taking into account possible changes in the $\Lambda_c/D$ ratio in
Au+Au collisions.

RHIC experiments have observed an enhancement of baryon production in
the intermediate transverse momentum region ($1.5 < p_T < 6$
GeV/c)~\cite{baryons}. At $p_T=3$~GeV/c, while in nucleon-nucleon
collisions, one proton is produced for every three pions (1:3), in
Au+Au collisions protons and pions at this $p_T$ are created in nearly
equal proportion (1:1). This enhancement of baryon production is seen
for protons and hyperons ($\Lambda$ (uds), $\Xi$ (dss) and $\Omega$
(sss)) along with their anti-particles~\cite{sorensenqm}. The
enhancement can be described by models involving multi-quark or gluon
dynamics during hadronization~\cite{reco}, models making use of baryon
junction loops~\cite{junctions}, or models with long range coherent
fields ({\it i.e.} strong color fields)~\cite{colorfields}. Until now,
however, the implications of possible charm-baryon enhancements on the
non-photonic electron spectra have not been considered.

The branching ratio for $\Lambda_c \rightarrow e$~+~\textit{anything}
(4.5\% $\pm$ 1.7\%) is smaller than that for $D^{\pm} \rightarrow
e$~+~\textit{anything} (17.2\% $\pm$ 1.9\%) or $D^0 \rightarrow
e$~+~\textit{anything} (6.87\% $\pm$ 0.28\%)~\cite{pdg}. In this case,
even if charm quark production is unchanged, increasing the
$\Lambda_c/D$ ratio will lead to a reduction in the number of observed
non-photonic electrons. In this report we assume that the
$\Lambda_c/D$ ratio is the same as the $\Lambda/K_S^0$
ratio~\cite{starratios}. Unless specified otherwise, the symbol $D$
represents the sum of $D^0$, $D^{+}$, and
$D_s$. PYTHIA~\cite{pythia} is used to generate the decay electron
spectrum from the input charm hadrons. We find that even when the
total charm hadron production in Au+Au collisions scales with the
number of binary nucleon-nucleon collisions $N_{bin}$ (\textit{i.e.}
total charm hadron $R_{AA}=1$~\cite{raanote}), the non-photonic
electron spectrum at intermediate $p_T$ can be supressed by as much as
20\%. We also present the non-photonic electron spectrum when the
total charm hadron $R_{AA}$ follows the measured charged hadron
$R_{AA}$~\cite{highptpapers}.  We find that if charm baryons are
enhanced as much as lighter flavor baryons, preliminary non-photonic
electron measurements imply a smaller suppression of charm quarks than
light quarks~\cite{nonphotonic}.

\begin{figure}[htbp]
\centering\mbox{
\includegraphics[width=0.5\textwidth]{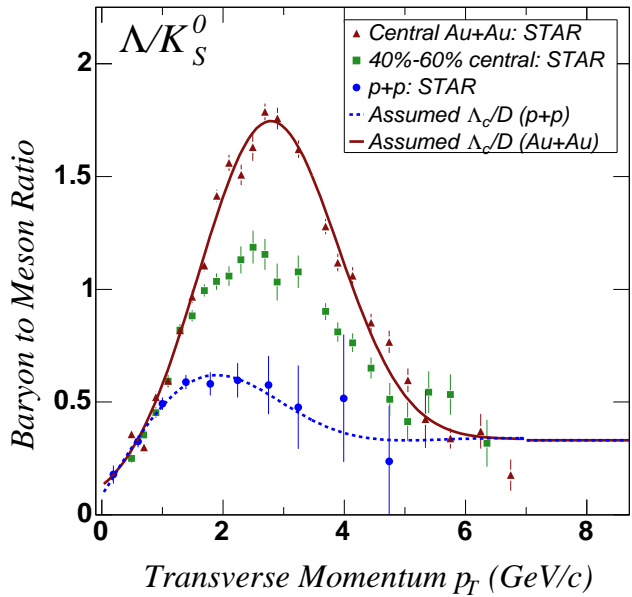}}
\caption{ (color online). $\Lambda/K^0_S$ ratio. The lines show the
  functional form of the the $\Lambda_c$/D-meson ratio used in our
  analysis. } \label{fig1}
\end{figure}

{\it Results.} --- Fig.~\ref{fig1} shows the $\Lambda/K^0_S$ ratio in
p+p and Au+Au collisions at
$\sqrt{s_{_{NN}}}=200$~GeV~\cite{starratios}. The $\Lambda/K^0_S$
ratio is larger than the $p/\pi$ ratio and the baryon enhancement
becomes even stronger for multi-strange baryons~\cite{msbraa}. In the
following analysis we take the $\Lambda_c/D$ ratio to have the same
form as the $\Lambda/K^0_S$ ratio. For $p_T >6.5$~GeV/c where the
$\Lambda/K^0_S$ ratio is unknown, we take the value $\Lambda_c/D =
0.33$. Since the source of the baryon enhancement at intermediate
$p_T$ is still under debate, it's difficult to assess the validity of
our assumed $\Lambda_c/D$ ratio. Possible explanations for the
enhancement include~\cite{sorensenqm} radial flow pushing heavy
baryons from lower $p_T$ into the intermediate $p_T$ region, baryon
junction dynamics, and enhanced production through coalesence or
recombination of quarks. We are not aware, however, of predictions for
the $p_T$ dependence of the $\Lambda_c/D$ ratio.

\begin{figure}[htb]
\centering\mbox{
\includegraphics[width=0.5\textwidth]{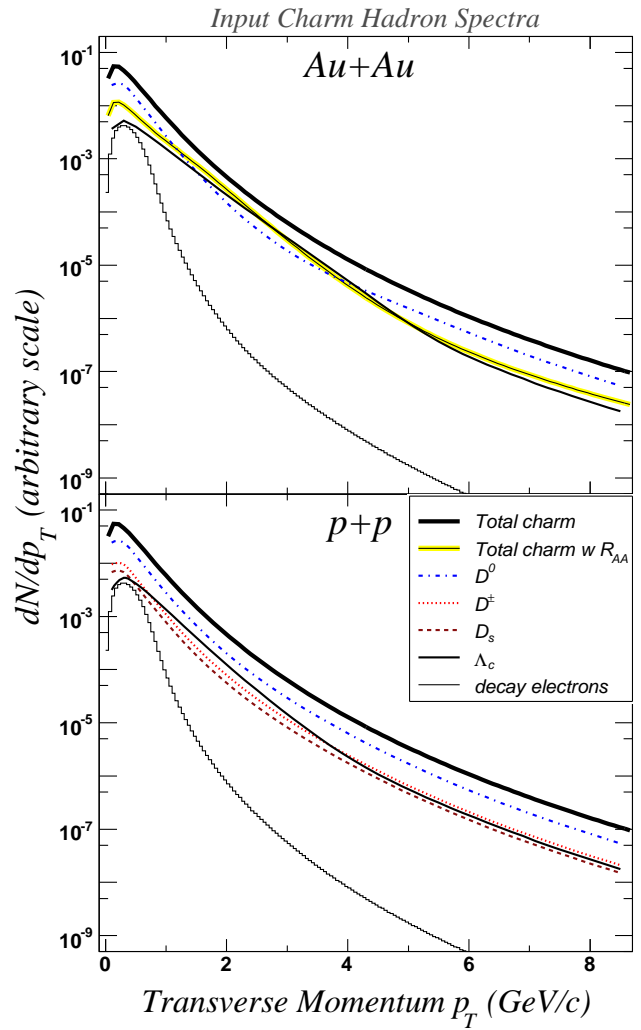}}
\caption{ (color online). The individual charm hadron and total charm
  hadron spectra. Here, the total charm hadron spectrum is assumed to
  follow a power-law shape with $\langle p_T \rangle = 1.3$~GeV/c and
  $n=9$. The individual charm hadron spectra are derived using the total
  charm spectrum and the assumed shape of the $\Lambda_c/D$ ratio in
  p+p collisions (bottom panel) or Au+Au collisions (top panel). We
  also show the total charm hadron spectra assuming a total charm
  hadron $R_{AA}$ similar to the measured charged hadron $R_{AA}$.
  The $D_s$ and $D^{\pm}$ spectra are omitted from the top panel for
  clarity. } \label{fig2}
\end{figure}

Fig.~\ref{fig2} shows the spectra for $D^0$, $D^{\pm}$, $D_{s}$, and
$\Lambda_{c}$. The spectra are derived such that the sum of the $D^0$,
$D^{\pm}$, $D_{s}$, and $\Lambda_{c}$ spectra follows a power-law, the
$\Lambda_{c}/D$ ratio has the form shown in Fig.~\ref{fig1}, and the
$D$-meson spectra all have the same $p_T$ dependence. Since we are
interested in the shape of the spectra, the scale of the y-axis is
arbitrary. The non-photonic electron spectrum will also be sensitive
to the $D^{\pm}/D^{0}$ and the $D_{s}/D^{0}$ ratios (the $D_s
\rightarrow e$~+~\textit{anything} branching ratio is
$8^{+6}_{-5}$\%~\cite{pdg}). An increase in the $D_{s}/D^{\pm}$ ratio
can therefore lead to fewer decay electrons depending on the poorly
known branching ratio.  At intemediate $p_T$, the $K/\pi$ ratio in
Au+Au collisions is enhanced compared to p+p
collisions~\cite{phpid}. One can also investigate how modifications to
the $D_s/D^{\pm}$ ratio in Au+Au collisions affect the non-photonic
electron spectrum. Since the enhancement in the $\Lambda/K^0_S$ ratio
is larger than the enhancement in the $K/\pi$ ratio, and since the
branching ratios for $D_s \rightarrow e$~+~\textit{anything} and $D^0
\rightarrow e$~+~\textit{anything} are similar, we expect a charm
baryon enhancement to have a larger effect on the decay electron
spectrum.  For this reason, in this report we use $p_T$ independent
relative $D$-meson abundances of 18, 7, and 5 for $D^0$, $D^{\pm}$,
and $D_{s}$ respectively~\cite{xinsthesis}.

\begin{figure}[htbp]
\centering\mbox{
\includegraphics[width=0.5\textwidth]{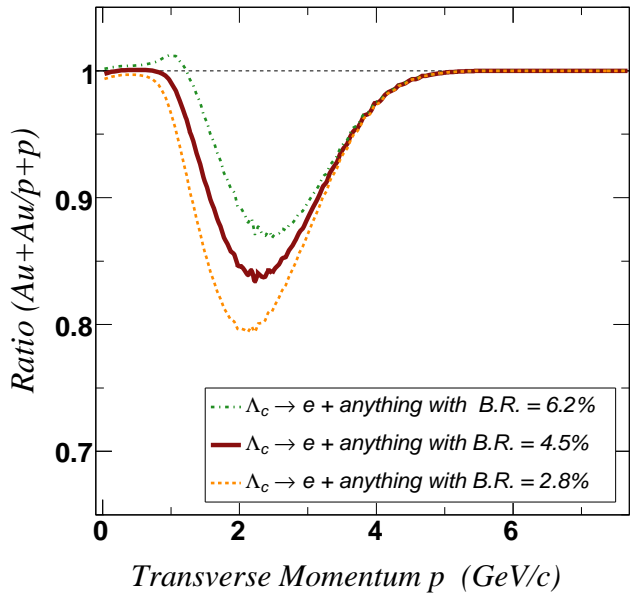}}
\caption{ (color online). Electron spectrum with $\Lambda_{c}$
  enhancement divided by the spectrum without $\Lambda_{c}$
  enhancement.  } \label{fig3}
\end{figure}

In Fig.~\ref{fig3} we show the effect of a $\Lambda_{c}$ enhancement
on the charm decay electron spectrum. The ratio of two cases is taken:
$\Lambda_{c}/D$ follows the shape of the $\Lambda/K^0_S$ ratio in
Au+Au collisions, or it follows the shape of the $\Lambda/K^0_S$ ratio
in p+p collisions. A supression of electrons from heavy flavor decays
due to the larger charm baryon-to-meson ratio in Au+Au collisions is
visible. The suppression in this figure is a result of smaller
$\Lambda_{c} \rightarrow e$~+~\textit{anything} branching ratio, which
has large uncertainties.  The highest and lowest curves show the cases
corresponding to the upper and lower experimental uncertainties on the
branching ratio~\cite{pdg}. The figure demonstrates that even if the
total charm yield follows $N_{bin}$ scaling, the non-photonic electron
spectrum may be suppressed. The magnitude of the suppression depends
on the $\Lambda_c/D$ ratio and the $\Lambda_{c} \rightarrow
e$~+~\textit{anything} branching ratio. The $\Lambda_c/D$ ratio in
Au+Au collisions is unknown but for the charm baryon-to-meson ratio
assumed here, the suppression can be as large as 20\%.

\begin{figure}[htbp]
\centering\mbox{
\includegraphics[width=0.5\textwidth]{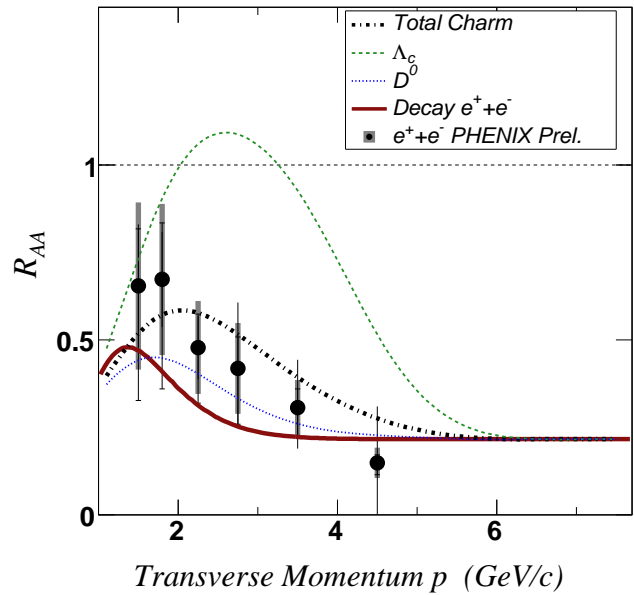}}
\caption{ (color online). $R_{AA}$ for charm hadrons and non-photonic
  electrons. The total charm spectrum in Au+Au collisions is scaled by
  the charged hadron $R_{AA}$ values. In this way the total charm
  hadron $R_{AA}$ has the same form as the charged hadron
  $R_{AA}$. The $\Lambda_{c}/D$ ratio is given the same form as the preliminary
  $\Lambda/K^0_S$ ratio. For $p_T<6$~GeV/c, the resulting decay electron
  $R_{AA}$ is smaller than either the $D$-meson or total charm
  $R_{AA}$. } \label{fig4}
\end{figure}

The presence of a charm baryon enhancement will change the charm quark
energy loss inferred from the preliminary non-photonic electron
$R_{AA}$ data. In Fig.~\ref{fig4} we show the case when the total
charm $R_{AA}$ has the same shape as charged hadron
$R_{AA}$~\cite{highptpapers}. In the lower $p_T$ region, this
assumption may not be realistic since the total charm quark production
is expected to follow $N_{bin}$ scaling. The error introduced,
however, will mostly affect the region below $p_T = 1.5$~GeV/c and may
be irrelevant to the higher $p_T$ regions of interest. Our analysis
shows that if the $\Lambda_{c}/D$ ratio has the form shown in
Fig.~\ref{fig1}, then the decay electron $R_{AA}$ at $p_T < 6$~GeV/c
will be smaller than the total charm $R_{AA}$.

In this report we have not considered contributions to the
non-photonic electrons from beauty decays. The $p_T$ value where the
yield of electrons from beauty decays is larger than from charm decays
is experimentally unknown. Theoretical calculations indicate that the
cross-over happens somewhere between $p_T =$ 3 GeV/c and $p_T =$ 10
GeV/c~\cite{bottom}. The branching ratios for beauty mesons and
baryons are not well know. We refer the reader to Ref.~\cite{bottom}
for discussion of the contribution of beauty to the non-photonic
electron spectrum.

In the intermediate $p_T$ region, the preliminary non-photonic
electron data are systematically above our calculations for the decay
electron $R_{AA}$~\cite{nonphotonic}. In the case that the heavy
flavor baryons have an enhancement similar to the light flavor
baryons, the non-photonic electron data indicate that the suppression
for charm quarks is smaller than that for light quarks. We varied the
input total charm hadron $R_{AA}$ and made a $\chi^2$ comparison to
the PHENIX data (with the systematic and statistical errors added in
quadrature). For $p_T > 2.0$~GeV/c, the PHENIX non-photonic electron
data are better represented when the total charm hadron $R_{AA}$ is
35\% greater than charged hadron $R_{AA}$. At $p_T$ near 6 GeV/c the
derived decay electron $R_{AA}$ matches the charged hadron $R_{AA}$
and the preliminary non-photonic electron $R_{AA}$ data reported in
Ref.~\cite{starelectron}. This {\it may} indicate that at $p_T =
6$~GeV/c (within the large errors) the total charm hadron suppression
is as large as the light hadron suppression. In light of the results
of this analysis, however, we believe one must also consider that a
charm baryon enhancement could extend to a higher $p_T$ than assumed
here. Direct measurements of heavy flavor hadrons are therefore needed
in order to accurately assess the energy loss of charm quarks.

{\it Summary.} --- We have studied the effect of the $\Lambda_c/D$
ratio on the non-photonic electron $R_{AA}$. We find that even when
the total charm hadron production scales with the number of binary
nucleon-nucleon collisions, an increase in the $\Lambda_c/D$ ratio
similar to that seen for the $\Lambda/K_S^0$ ratio can lead to a
suppression in central Au+Au collisions of non-photonic electrons at
intermediate $p_T$. This may help explain why the non-photonic
electron $R_{AA}$ is smaller than was predicted by radiative energy
loss models: models which are able to describe the light hadron
$R_{AA}$.  If the $\Lambda_c/D$ ratio has the form assumed in this
report, the PHENIX non-photonic electron data at intermediate $p_T$
prefer a total charm hadron $R_{AA}$ 35\% larger than the charged
hadron $R_{AA}$ --- implying less energy loss for charm quarks than
light quarks. If the relative fractions of charm hadrons are not
altered in Au+Au collisions compared to p+p collisions, the
non-photonic electron $R_{AA}$ values are difficult to understand
within current radiative energy loss models. Since the non-photonic
electron measurements depend on the $D^{0}/D$, $D^{\pm}/D$, $D_s/D$
and the $\Lambda_c/D$ ratio, direct measurements of heavy-flavor
hadron yields are needed to draw firm conclusions regarding energy
loss for heavy quarks. These measurements will only be possible at
RHIC with detector upgrades~\cite{upgrades}.


\par {\bf Acknowledgements}: We thank N. Xu for his comments and
suggestions which have greatly improved this work. One of us
(P.~R.~S.)  would like to thank the Battelle Memorial Institute and
Stony Brook University for support in the form of the Gertrude and
Maurice Goldhaber Distinguished Fellowship. This work was supported in
part by the National Natural Science Foundation of China under grant
number 10475071.



\begin{thebibliography}{00}
\bibitem{raanote} $R_{AA}$ is the ratio of the $p_T$ spectrum in
  nucleus-nucleus collisions and p+p collisions where each has been
  scaled by $N_{bin}$ to account for the trivial increase in particle
  production with collision system-size.
\end{thebibliography}
\end{document}